# Exploring patterns in European singles charts

Andrzej Buda[1,2], Andrzej Jarynowski[3,4]

[1]Wydawnictwo Niezależne, Wrocław, Poland
[2] IFJ, Polish Academy of Science, Cracow, Poland
[3]Jagiellonian University, Institute of Physics, Cracow, Poland
[4]Moldova State University, Kishinev, R. of Moldova

andrzej.jarynowski@sociology.su.se, andrzejbudda@gmail.com

**Abstract.** European singles charts are important part of the music industry responsible for creating popularity of songs. After modeling and exploring dynamics of global album sales in previous papers, we investigate patterns of hit singles popularity according to all data (1966-2015) from weekly charts (polls) in 12 Western European countries. The dynamics of building popularity in various national charts is more than the economy because it depends on spread of information. In our research we have shown how countries may be affected by their neighbourhood and influenced by technological era. We have also computed correlations with geographical and cultural distances between countries in analog, digital and Internet era. We have shown that time delay between the single premiere and the peak of popularity has become shorter under the influence of technology and the popularity of songs depends on geographical distances in analog (1966-1987) and Internet (2004-2015) era. On the other hand, cultural distances between nations have influenced the peaks of popularity, but in the Compact Disc era only (1988-2003). We have also indicated the European countries "in line" with global trends e.g. The Netherlands, the United Kingdom and outsiders like Italy and Spain.

## Introduction

In academic studies, as well as real-world settings (business-oriented or fan-oriented) many metrics of measuring popularity [1] and distance were proposed [2]. Computer science and mathematics introduced a plenty of ranking algorithms as Google's PageRank of WWW sites and its variation.

Music – mediated by sounds – is widely enjoyed by people all over the planet, across a diverse background of numerous cultures. Some common bonds of music are dynamics, texture, rhythm and pitch. While music have accompanied humans on their journeys through life, and the people themselves, in turn, the created of the musical trends and established businesses to supply demand on music.

Recently, the knowledge of computational tools for sociology and music such as recommendations [3] has undergone an accelerating growth, however all models of such system are incomplete without real data, especially register-based [4]. The requirement to calibrate the algorithms encourages cooperation between various registering institutions, which, in turn, exerts a pressure on collecting data for simple analysis by many researchers who work on new models and use complex tools often taken from other disciplines [5]. We present computational approach to this issue. We based our project on different approach from beginning with mathematical though computer science to finish with sociological and statistical analyses. This work is also very specious, because of its interdisciplinary and showing light on methods which were not use in "music studies" till this moment. Only a few researches were provided for understanding successes or failures of an artist, mainly in economic frameworks of product life-cycle or innovation spread. There were also culture studies focused on ethnology and customers decisions. Unfortunately, quantitative methods are rarely used in these analysis. If it is applied, linear regression dominates. A slightly less well known but related superstar phenomenon is even more interesting from complex system perspective due to herd effect and criticality of customers decision. Music has had an impact on society for as long as people have made music and listened to it. Music has acted as a means for giving a cultural voice to people – consumers. Music even settles communities: for instance, punk, techno, and ethnic music all embody some type of subculture that encompasses an entire lifestyle. In this work, we model the challenges for artists and their labels staying behind their market decisions in ultimate sales game. The properties of phonographic market have been recently explored from economic point of view including the stochastic model of global phonographic market and the influence of digital piracy on albums sales dynamics [6]. Like in financial markets, according to album sales history, it is possible to use Minimum Spanning Trees methods (MST) and distinguish sectors and subsectors of artists that represent various music genres. However, that kind of hierarchical structure does not contain sectors that represent pop music, because there are two groups of customers: music fans that buy records every week, and occasional clients who spend their money on music before Christmas or Valentine's Day [7]. This second group is only interested in celebrities from various music genres and makes record sales grow up at the end of each year. According to these economic reasons, pop music does not exist in hierarchical structure of the market [6], but in the Minimum Spanning Tree we have the sector of the most famous celebrities that represent various genres. Thus, the group of occasional customers is responsible for seasonality that affects the global dynamics of albums sales.

On the other hand, there is also another part of phonographic market - the hit singles. Since Thomas Edison's phonograph, the short cuts of music has always been sold each year in form of 7" vinyl singles, 3" or 5" CD singles, mp3, ringtones, radio or TV broadcast, etc. The variety of forms is much bigger that physical long-playing records format, so the dynamics of hit singles popularity should be considered from information theory/spread and computational sciences point of view. Because each national singles chart is based on different rules. For example, the UK hit singles chart depends on physical or digital sales only, but the Italian singles chart is a combination of singles sales and airplays from the local radio and TV. The national singles markets have various sizes, so the sales results are usually published weekly. However, in the sixties and seventies the singles charts in Austria and France have been published monthly and the West German chart has been published bi-weekly.

There has been a lot of research into epidemiologic or opinions spread models [8], especially based on diffusions [9] or interactions between two nearest neighbours (the Ising model). The dynamics of hit singles popularity has such properties because one European nation might affect or have been affected by others in their neighbourhood.

Diffusion of information and influence spread within cultural sphere are also topics of broad evaluation. We have studied the diffusion of popularity of singles [10] in Western Europe in order to compare time delays of topping in the national charts. Countries are constantly influencing their surroundings and being influenced by others. The topological characteristics of networks consequently determine dynamical processes on top of the network, e.g. cascade of information adoption or default contagion in culture networks [11-14]. It gives us an opportunity to observe the spread of music popularity.

## Data analysis and background information

In our research we have collected all the hit singles data from the last 50 years. We have selected the most popular singles in the period 1966-2015 that have reached the weekly published Top 15 in the national hit singles charts including: 01 AUSTRIA (A) 02 BELGIUM (B) 03 SWITZERLAND (CH) 04 GERMANY (D) 05 SPAIN (E) 06 FRANCE (F) 07 GREAT BRITAIN (GB) 08 ITALY (I) 09 IRELAND (IRL) 10 NORWAY (N) 11 NETHERLANDS (NL) 12 SWEDEN (S)

These data have been published weekly on European music papers, radio or TV broadcast and the Internet for the last 50 years. Of course, since 1966 we have had the physical analog formats of music (7 inch singles, 12 inch maxi singles), digital 3 or 5 inch singles that have been dominated the charts in the nineties (1988-2001) and the streaming of official digital mp3, ringtones (2002 until now). We have also investigated how the spread of hit singles through various European countries may depend on these formats of music dominating various ages.

In our research we have considered trajectories of a hit single performed on national charts as a product lifecycle with the highest position on each chart. If the highest position belongs to the Top 15, the country has been affected. Otherwise, the single has not been a hit, so the country is not affected. As a main data parameter $t$ we consider a number of weeks (after initial single release) when the single reaches the peak of the national record chart. If the single does not affect the country (or has a peak outside the Top 15), we set the value of this parameter $t = 52$ weeks, because in phonography the maximum length of seasonal popularity is usually limited to one year.

We have analysed all period (1966-2015) where initial singles releases have been more or less contagious and come from various countries. Usually, the hit singles popularity have come from the neighbourhood countries. According to our results, the United Kingdom, the Netherlands and Belgium have ability to be affected fast, after a few weeks (Fig. 1). On the other hand, Spain and Italy have been affected after $t = 20$ weeks after initial hit single release. This kind of behaviour suggests geographical dependences, but in our research we focus on various eras (1966-1987, 1988-2003, 2004-2015) and consider also cultural dependences between European countries.

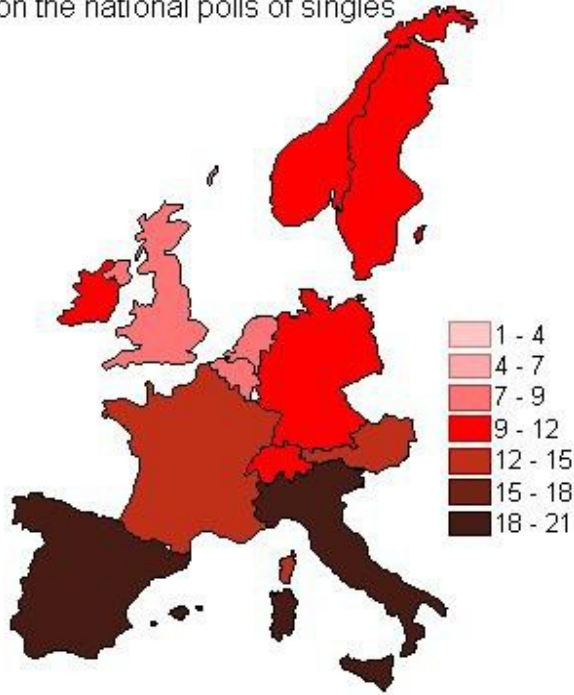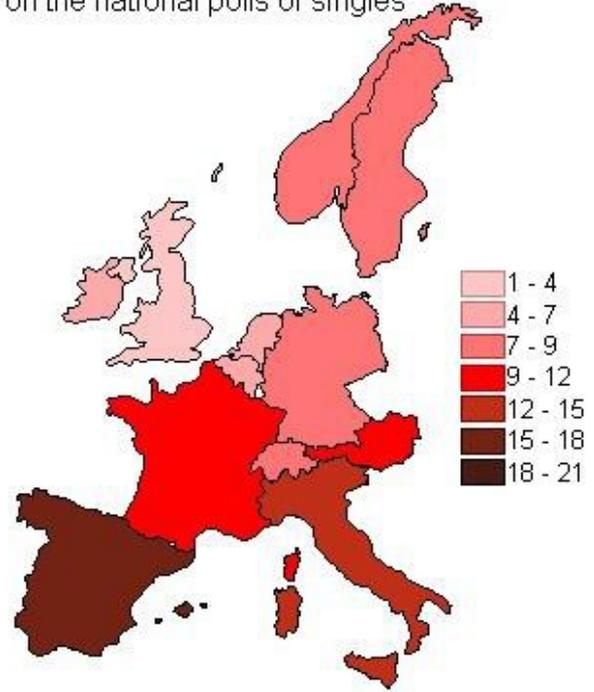

**Fig. 1** The map of western Europe according to spread of music popularity for hit singles (1966-2015)

We investigate distances between European countries in Cartesian / Euclidean two dimensional meaning. The most typical representation are geographical distances. However, many other matrixes of distances are possible. For example cultural map of Europe was created by political scientists based on the World Values Survey [15]. Two dominant dimensions were selected (explaining 70% of variations between countries): traditional versus secular-rational values on the vertical y-axis and survival versus self-expression values on the horizontal x-axis (Fig. 2).

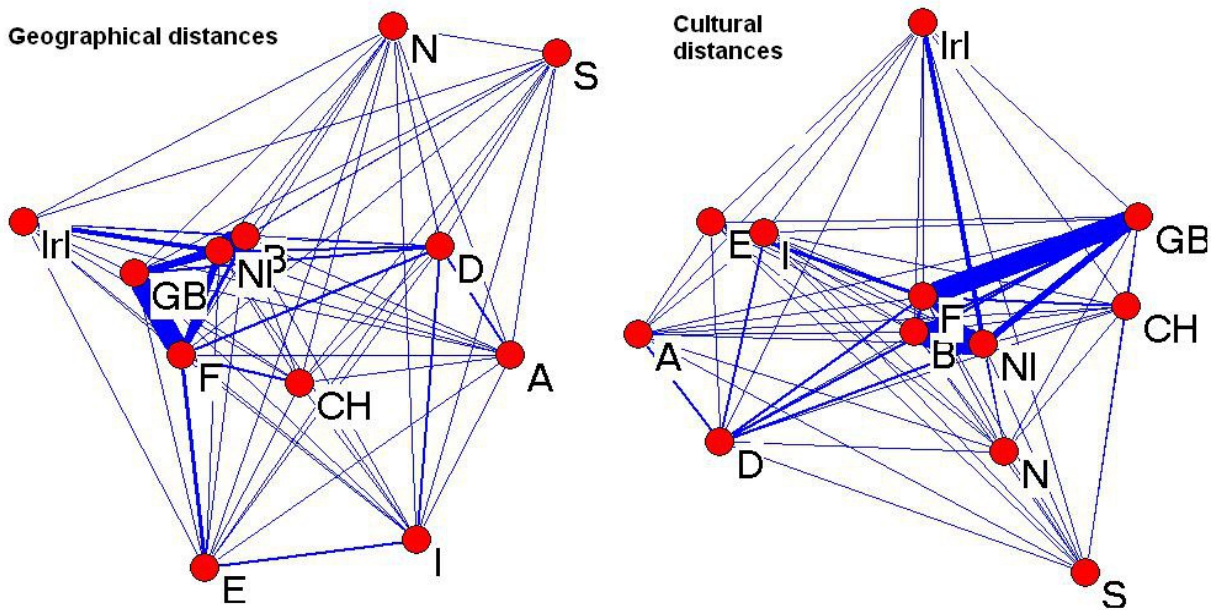

**Fig. 2** The map of western Europe according to geographic and cultural distance. Links are based on flows between counties (inverse gravity force)

In physical and digital era (LP, CD) the product life-cycle had its maximum a few weeks after the release, because music labels need more time to supply physical phonograms and satisfy the number of records demanded by customers. On the other hand, in recent Internet era, *t* usually tends to 1, because records may reach the peak of popularity in the first week after the release. The example of this behavior in global record sales trajectories of Pink Floyd (1979) and Coldplay (2007) is given by (Fig. 3).

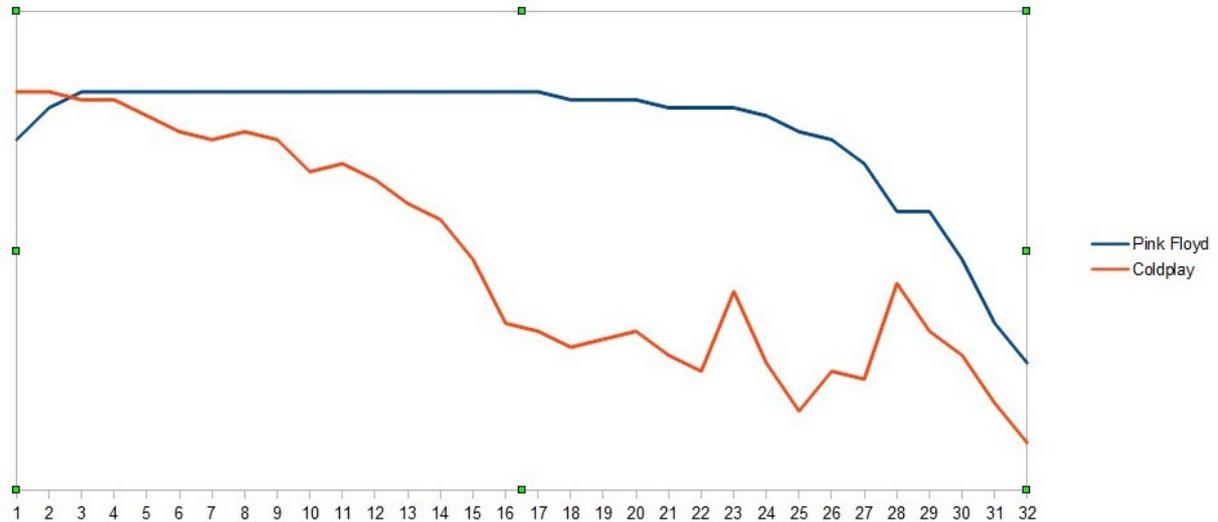

**Fig. 3** The position of The Wall by Pink Floyd (1979, upper line) and Viva La Vida by Coldplay (2007) on global albums sales chart in consecutive weeks after the release.

However, the patterns in local European singles charts mean more than just the economy, because all countries might be affected by their geographical neighbourhood (Fig 1). Moreover, the cultural distances have also influenced on information spread, and peaks of popularity. Thus, we have decided to show normalized distances between countries according to geography, culture and *t* - time delay between peaks of popularity after singles premieres (Fig. 4). The double cross on the third diagram (Fig. 3) reflects the fact that Spain and Italy has become the most isolated countries according to spread of hit singles popularity in Europe.

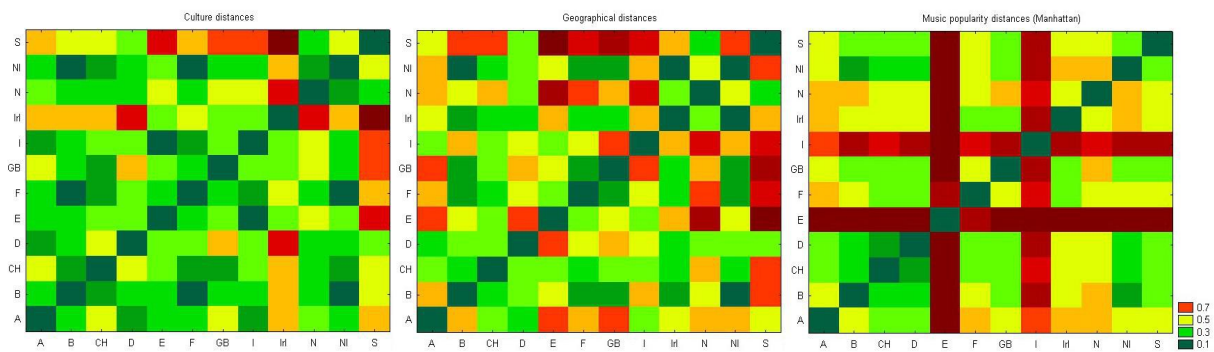

**Fig. 4** Normalized distances between countries (in geography, culture and music popularity)

## Results – Phases.

In our research we explore patterns in European singles chart in long time period (1966-2015) where the time delay *t* between record premiere and the peak of popularity has become shorter and shorter. This phenomena refers to singles and albums (Fig. 2). So it is reasonable to distinguish analog (1966-1987), digital (1988-2003) and Internet (2004-2015) era in record industry, because of the technology.

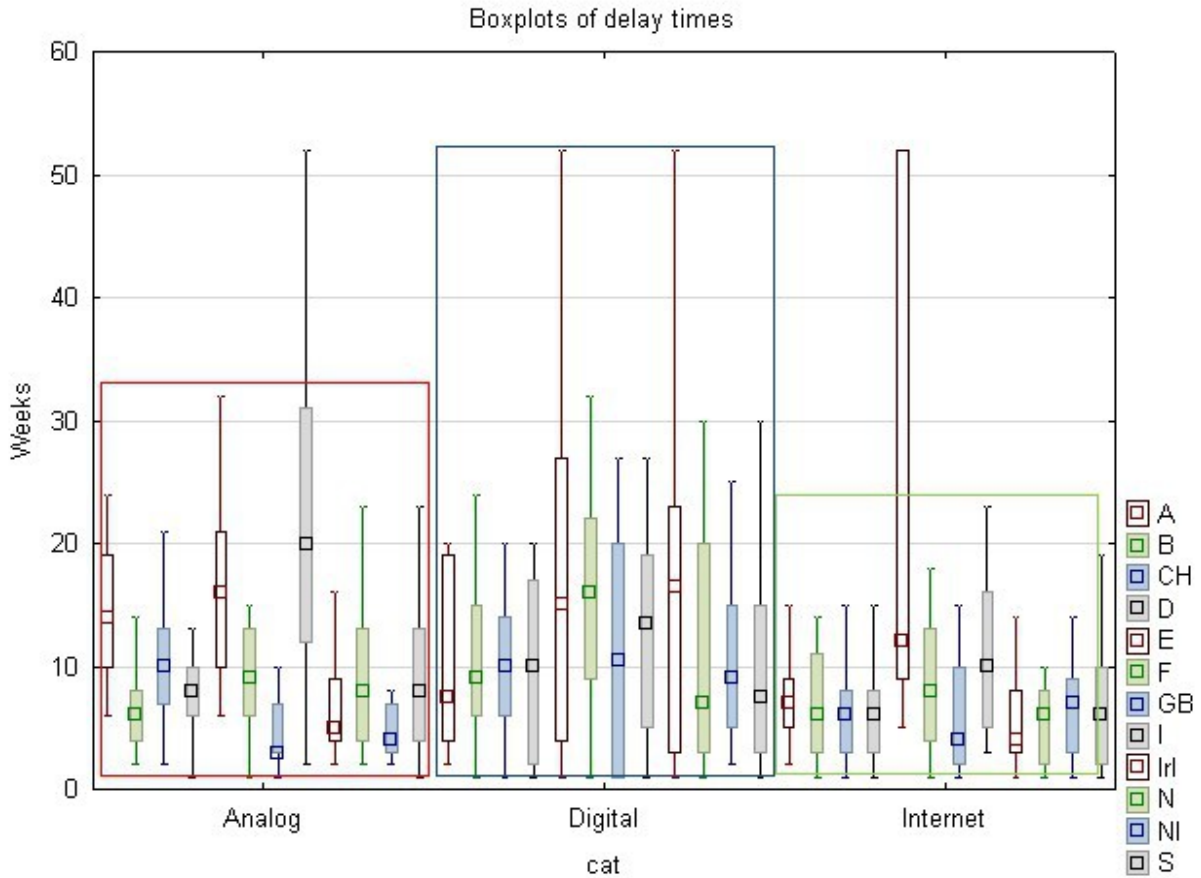

**Fig. 5** Boxplots of distrubuitions of delays in peeaking top position on charts in analog, digital and Internet era.

1) Analog era (1966-1987)

In the early '50s of the 20[th] century, as a physical format, the 7 inch singles were the simpliest and basic formats of songs that make record industry able to compile national record sales charts. We have observed geographical dependences between popularity spreads. Of course, most of hit singles have initial roots in the UK and gained global popularity quickly, but we also considered songs from all 12 European countries that had national charts on that time. If the hit single came from distant countries like Norway or Sweden, it took half or one year to affect all Europe. For example, Take On Me by Norwegian trio A-ha gained the peak of popularity and effected all 12 European countries in the late 1985, one year after initial release. Thus, the correlation between $t$ and geographical distance [Tab. 1] is observed for all countries (Fig. 5).

2) Digital era (1988-2003)

The '80s of the 20[th] century introduced the „undestructible" digitality to music industry in physical forms and Compact Discs dominated phonographic markets all over the world. In digital the peak of national charts had been reached almost equally in all countries. Moreover, we observe significant correlations between $t$ and cultural distances in Europe [Fig. 4, Tab. 1]. On the other hand, the end of 20[th] century brought almost all phonographic markets into European integration and unity. So, even the Italy, that had never created an international hit in 7" singles analog era (1966-1987), had been finally able to initialize the popularity of Dragostea Din Tei by Moldovian band O-Zone (2003). The language of this song was understamble to imigrants from Romania that had been able to move and settle in other European countries in the early 21[th] century. The dependence between music popularity and geographical distances disappeared in that era [Tab. 1].

3) Internet Era (2004-2015).

The growing power of Internet in the beginning of 21[th] century caused dramatic decrease of albums sales all over the world [16] because people could have direct access to music files by Internet connection. There was no need to collect music in traditional physical formats like CDs or vinyls because of mp3s and ringtones. On the other hand, music labels developed streaming technologies to stop digital piracy. Thus, all national charts had become compiled also from digital singles sales. The Internet era caused globalization in the phonographic markets all over the world (2004-2015). Thus, the time $t$ of reaching peaks of popularity had become much shorter than before and correlations between $t$ and culture distances had disappeared. On the other hand, we could observe weak correlations between $t$ and geographical distances [Tab. 1] because all effects of the Internet era mean stronger dependence on information spread.

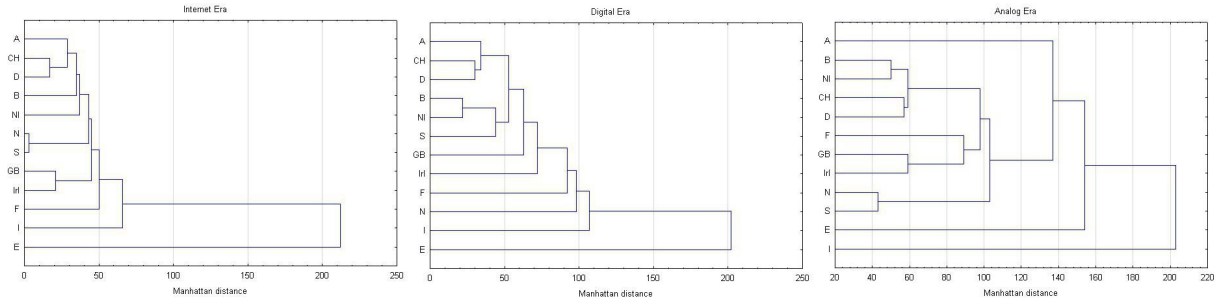

**Fig. 6** Hierarchical representations of delays $t$ in peaking top position on charts in analog, digital an Internet era

Subdominant ultrametric space that provides hierarchy of dependence in other perspective like diagrams (Fig. 6). This method describes collective relations between elements [17] – countries in Europe. We have also computed the Manhattan distances between peaks of popularity in 12-dimensional space represented by 12 national charts. Thus, it is possible to detect hierarchical structure of European singles chart in ultrametric space. It has been necessary to show the evolution in hierarchical structure of European countries. The United Kingdom has always been a leader in spreading popularity of songs. The most distant country from the rest is Italy, because a large number of European hit singles (including Dancing Queen by Abba or Macarena by Los Del Rio) has never gained popularity there. However, this phenomena of isolation has disappeared completely because of the Internet era. On the other hand, another distant country – Spain – has always been affected late because of geographical reasons. Moreover, Spain is going to be more isolated than before, because a number of 21$^{st}$ century popular hits has never reached Spanish charts even in the Internet era.

**Tab. 1** Correlations between networks of distances between European countries. Music delays categories 1- Analog, 2- Digital, 3- Internet, where med - median and mean – mean delays. Geo – geographical and cult – cultural distance stand for reference.

| Corr. | 1mean | 1med | 2mean | 2med | 3mean | 3med | geo | cult |
|---|---|---|---|---|---|---|---|---|
| geo | 0.559921 | 0.631344 | 0.124026 | -0.282997 | 0.236335 | 0.338796 | 1.000000 | -0.456454 |
| cult | -0.122683 | -0.316936 | 0.260181 | 0.503714 | -0.086489 | -0.328758 | -0.456454 | 1.000000 |
|  | Analog | Analog | Digital | Digital | Internet | Internet |  |  |

## Conclusions and future works.

Diversity of gaining popularity in European countries is significant and has become strongly investigated in the Internet era in spite of globalization [18]. The whole history of record industry shows that music labels tend to supply brand new records to customers as fast as it is possible. Thus, in the next consecutive technological era, time delays $t$ between the hit single release and the height of popularity become shorter and shorter (Fig. 5, 6). Thus, the network and information society has been driven by the technological forces. However, each technological era has its own patterns with different possibility of modelling it [19]:
- Analog regime: huge variety between countries and medium variety within countries (Fig. 5), positive spatial correlations [Tab. 1]. The nationality of all considered singles charts in Europe has meaning. Geography plays a main role in spread of popularity (grid modelling approach could describe that era well).
- Digital regime: huge variety inside the countries, but small differences between countries (Fig. 5), positive correlations with culture distances [Tab. 1]. It could be the effect of European integration and unification (decreasing distance between countries). The variety of $t$ within each country is much bigger, probably because of new music genres. Here, cultural similarities have the strongest impact on information spread. To model this system in proper way, networks approach is needed.
- Internet era: medium variety between countries and medium variety within country (Fig. 5), no correlations significant [Tab. 1]. Despite of globalization, local nationalism come back on the stage. Here, modelling would be more difficult, because none of presented topologies (Fig. 2) may describe information spread well.

Results of our visualization analysis may develop new (more reliable) dynamic description music popularity spread patterns. Some temporal [20] as well as spatial [21] relations were already known to scientific community, but not from singles popularity perspective. However, our results are limited to 12 available European singles charts, so the effect of isolation in Spanish singles charts in the Internet era remains unexplained. If this phenomena is caused by globalisation, it might be probably successfully explained by considering data and influences from other non-European countries including Latin America. On the other hand, from the non-European point of view, the leadership of the United Kingdom might be the reminiscence of the United States and other English-speaking countries. To answer these questions, we are going to apply agent-based modelling to global phonographic market in our future works [19].

# Appendix

**Tab. 2** List of the 50 most popular singles in 1966-2015:

| Analog | Digital | Internet |
| --- | --- | --- |
| SCOTT MCKENZIE - SAN FRANCISCO | EDDY GRANT - GIMME HOPE JO'ANNA | CRAZY FROG - AXEL F |
| THE BEATLES - HEY JUDE | MARIAH CAREY - WITHOUT YOU | SHAKIRA/WYCLEFF JEAN - HIPS DON'T LIE |
| SERGE GAINSBOURG/JANE BIRKIN - JE T'AIME...MOI NON PLUS | SCATMAN JOHN - SCATMAN | GNARLS BARKLEY - CRAZY |
| MUNGO JERRY - IN THE SUMMERTIME | LOS DEL RIO - MACARENA | RIHANNA/JAY-Z - UMBRELLA |
| GEORGE HARRISON - MY SWEET LORD | ELTON JOHN - CANDLE IN THE WIND | TIMBALAND/ONE REPUBLIC - APOLOGIZE |
| VICKY LEANDROS - APRES TOI | CELINE DION - MY HEART WILL GO ON | DUFFY - MERCY |
| ROLLING STONES - ANGIE | CHER - BELIEVE | BEYONCE - IF I WERE A BOY |
| GEORGE MCCRAE - ROCK YOUR BABY | LOU BEGA - MAMBO NO.5 | LADY GAGA - POKER FACE |
| GEORGE BAKER SELECTION - PALOMA BLANCA | BOMFUNK MC'S - FREESTYLER | ADELE - ROLLING IN THE DEEP |
| ABBA - DANCING QUEEN | KYLIE MINOGUE - CAN'T GET YOU OUT OF MY HEAD | RIHANNA/CALVIN HARRIS - WE FOUND LOVE |
| BONEY M - MA BAKER | SHAKIRA - WHENEVER WHEREVER | PSY - GANGNAM STYLE |
| J.TRAVOLTA/OLIVIA N.JOHN - YOU'RE THE ONE THAT I WANT | LAS KETCHUP - ASEREJE | DAFT PUNK/PHARRELL WILLIAMS - GET LUCKY |
| VILLAGE PEOPLE - YMCA | EMINEM - LOSE YOURSELF | PHARRELL WILLIAMS - HAPPY |
| PINK FLOYD - ANOTHER BRICK IN THE WALL | O-ZONE - DRAGOSTEA DIN TEI | LILLYWOOD/THE PRICK - PRAYER INC. |
| BARBRA STREISAND - WOMAN IN LOVE | | MARK RONSON/BRUNO MARS - UPTOWN FUNK |
| PAUL MCCARTNEY/STEVIE WONDER - EBONY AND IVORY | | |
| DAVID BOWIE - LET'S DANCE | | |
| STEVIE WONDER - I JUST CALLED TO SAY I LOVE YOU | | |
| A.HA - TAKE ON ME | | |
| EUROPE - THE FINAL COUNTDOWN | | |
| MICHAEL JACKSON - I JUST CAN'T STOP LOVING YOU | | |